\documentclass[useAMS,usenatbib]{mn2e}

\usepackage{amsmath}
\usepackage{aas_macros}
\usepackage[T1]{fontenc}
\usepackage{pslatex}

\title{A magnetic accretion switch in pre-cataclysmic binaries?}

\author[Jeremy J.~Drake, Cecilia Garraffo, Dai Takei and Boris Gaensicke]{Jeremy J.~Drake$^1$,\thanks{E-mail:jdrake@cfa.harvard.edu} Cecilia Garraffo$^1$, Dai Takei$^1$ and  Boris Gaensicke$^2$\\
$^1$Harvard-Smithsonian Center for Astrophysics,
    60 Garden Street, Cambridge, MA 02138\\
$^2$Department of Physics, University of Warwick, Coventry CV4
  7AL, UK}

\begin{document}
\date{}
\pagerange{\pageref{firstpage}--\pageref{lastpage}} \pubyear{2013}

\maketitle

\begin{abstract}
We have investigated the mass accretion rate implied by published surface abundances of Si and C in the white dwarf component of the 3.62~hr period pre-cataclysmic binary and planet host candidate QS~Vir (DA+M2-4).   Diffusion timescales for gravitational settling imply $\dot{M} \sim 10^{-16}M_\odot$~yr$^{-1}$ for the 1999 epoch of the observations, which is three orders of magnitude lower than measured from a 2006 {\it XMM-Newton} observation.  This is the first time that large accretion rate variations have been seen in a detached pre-CV.  A third body in a 14~yr eccentric orbit suggested in a recent eclipse timing study is too distant to perturb the central binary sufficiently to influence accretion.  A hypothetical coronal mass ejection just prior to the {\it XMM-Newton} observation might explain the higher accretion rate, but the implied size and frequency of such events appear too great.   We suggest accretion is most likely modulated by a magnetic cycle on the secondary acting as a wind ``accretion switch'', a mechanism that can be tested by X-ray and ultraviolet monitoring.  If so,  QS~Vir and similar pre-CVs could provide powerful insights into hitherto inscrutable cataclysmic variable and M dwarf magnetospheres, and mass and angular momentum loss rates.
\end{abstract}




\section{Introduction}

Short-period binaries composed of a white dwarf and late-type star are
of fundamental importance to astrophysics and are the progenitors of
cataclysmic variables (CVs) and novae, some of which likely evolve to
form Type~1a supernovae.  
They are the outcome of a common envelope evolutionary phase in
which friction leads to rapid orbital shrinkage of an
initially wider binary \citep{Paczynski:76}.   The timescale for initiation of mass transfer 
and the subsequent orbital evolution of a CV depend critically on the angular momentum loss (AML) rate.   For orbital periods above the CV period gap of $\ga 3$~hr, AML is dominated by the magnetized wind of the M dwarf secondary.
While AML through gravitational radiation, which is significant for periods $\la 3$~hr, is theoretically well-understood for the stars in CVs, there is no comprehensive theory of spin-down through magnetized winds.  
AML depends on the mass loss rate and the large-scale stellar magnetic field \citep[e.g.][]{Weber.Davis:67,Mestel:68,Kawaler:88}.   Mass loss rates through winds and coronal mass ejections (CMEs) remain extremely difficult to measure for late-type stars and are especially uncertain at the very rapid rotation rates of close binaries where magnetic proxies such as X-ray emission show saturation effects.  Plausible values lie in the range $10^{-15}$--$10^{-12} M_\odot$~yr$^{-1}$ \citep[see, e.g., the summary by][]{Matranga.etal:12}.  Consequently, mass loss rates in AML prescriptions used for models of CV evolution are based largely on guesswork or simple extrapolations from the solar case, and can differ by orders of magnitude.  \citet{Knigge.etal:11} have highlighted the  enormous range in predicted AML loss, even for conceptually similar AML prescriptions.  

The attendant complexity of mass transfer and an accretion disk, and the resulting smothering of signals from the M dwarf secondary itself, renders any direct studies of CV secondary mass loss and AML implausible---one reason current AML prescriptions are so uncertain in a class of stars that has been studied in detail for decades.  Pre-cataclysmic binaries close to contact  but unencumbered with all the complications and emission associated with Roche Lobe overflow accretion, offer much more direct ways to study winds and magnetospheres of M~dwarf secondaries with CV-like rotation periods and magnetic activity.  This class of object also exhibit orbital period variations that appear to defy explanation, but that might be related to magnetic activity.

The intriguing white dwarf--M dwarf eclipsing binary  QS Vir (DA+M2-4 3.6~hr; formerly known as EC~13471$-1258$) was first discovered in the Edinburgh-Cape blue object survey \citep{Kilkenny.etal:97}.  It has gained considerable recent attention as a potential diagnostic of close binary evolution and angular momentum loss in cataclysmic variables, as well as controversy regarding currently unexplained orbital period variations.  High-speed and multi-colour photometry, together with UV
HST STIS and visible light spectroscopy, initially suggested the system just filled its Roche Lobe 
\citep{Kawka.etal:02,O'Donoghue.etal:03}.
Subsequent studies favoured a detached system but found evidence for material possibly associated with wind accretion and prominences from the secondary within the Roche lobe of the white dwarf \citep{Ribeiro.etal:10,Parsons.etal:10}.

The orbital period of QS~Vir has presented puzzling variations.  \citet{O'Donoghue.etal:03} noted ``jitter" of up to 12~s in the eclipse timings they attributed to 
magnetic cycling in the M dwarf.  With observations over a slightly longer baseline, 
\citet{Qian.etal:10} inferred the presence of a giant planet with mass $6.4 M_{Jupiter}$ in a 7.86~yr orbit.  \citet{Parsons.etal:10} ruled this specific set of parameters out from more complete eclipse monitoring that revealed a 150~s drop in eclipse times in later epochs.  They found a third body in a highly elliptical orbit best fits the more complete data, noting that this explanation for the secular orbit change is also not without problems.  \citet{Almeida.Jablonski:11} favoured a two-companion solution to the orbital variations, although \citet{Horner.etal:13} have recently shown all proposed planetary solutions to the problem to be dynamically unstable.  Similar orbital period variations also characterize several other close binary stars---9 out of 10 systems studied according to \citet{Zorotovic.Schreiber:13}, who argue that if planets are responsible, these are ``second generation", formed after common envelope evolution.   An alternative possibility is that magnetic cycles are responsible via the \citet{Applegate:92} or a related mechanism, although for QS~Vir the energetic requirements appear to be an order of magnitude too large \citep{Qian.etal:10,Parsons.etal:10}.

\citet{Matranga.etal:12} recently discovered deep, narrow primary X-ray eclipses in an {\it XMM-Newton} observation of QS~Vir obtained in 2006.  The eclipses revealed that the white dwarf dominated the emission rather than the M~dwarf, as would be expected for a detached system.  The low-resolution EPIC PN spectrum was well-matched by a cooling flow model with an accretion rate of $\dot{M} = 1.7 \times 10^{-13} M_\odot$~yr$^{-1}$ \citep[see also][]{Bilikova.etal:10}.  
\citet{Matranga.etal:12} speculated that the observed mass transfer could be through a wind, perhaps supplemented by upward chromospheric flows on the M dwarf, analogous to spicules and mottles on the Sun.  QS~Vir would then be on the very verge of Roche lobe overflow in a rare evolutionary phase lasting only of the order of a million years, making it an unlikely find. 

The \citet{O'Donoghue.etal:03} study included an analysis of HST STIS observations of QS~Vir to measure the white dwarf effective temperature, surface gravity and the system radial velocity curve.  As a product of that analysis, they estimated metal abundances.  Since metals are expected to settle out of the atmosphere of cool DA white dwarfs on relatively short timescales, the metal abundance can be used to estimate the mass accretion rate.  Here, we examine the implications of the abundances found by \citet{O'Donoghue.etal:03}.

\section{The white dwarf metal abundances}
\label{s:obs}



\citet{O'Donoghue.etal:03}  obtained UV STIS spectra of QS~Vir on 1999 August 28.  
At UV wavelengths, QS~Vir is completely dominated by the white dwarf component.  \citet{O'Donoghue.etal:03} performed a model atmosphere analysis and found an effective temperature $T_{eff} = 14 220$~K and surface gravity $\log g = 8.34$.  Additionally, they identified a number of lines of C~I, C~II and Si~II and used these to derive abundances of approximately 1/30 solar for C and 1/60 solar for Si.  Since metals gravitationally settle out of the photosphere of a cool white dwarf, the observed metal abundance is related to the mass accretion rate as follows \citep[e.g.][]{Dupuis.etal:93,Koester:09}
\begin{equation}
\label{e:mdot}
\dot{M}=\frac{q M_{WD}}{\tau_X} \frac{n(X_{WD})}{n(X_{RD})}.
\end{equation}
Here, $q$ is the mass fraction of the surface convection zone on the white dwarf within which accreted material is mixed, $\tau_X$ is the diffusion timescale for gravitational settling of species $X$ out of this zone, and $n(X_{WD})$ and $n(X_{RD})$ are 
the abundances by number in the white dwarf atmosphere and in material accreted from the red dwarf, respectively.

Using the white dwarf mass from \citet{O'Donoghue.etal:03}, the listings of the convection zone mass fraction, $q$, and diffusion timescales for C and Si in Tables~1 and 4 of \citet{Koester:09}, and assuming solar abundances for the secondary red dwarf as indicated by the \citet{O'Donoghue.etal:03} study, we find for a temperature $T_{eff}=14000\,$K mass accretion rates of $1.0\times 10^{-16}$ and $7.5\times 10^{-17}$~$M_\odot$~yr$^{-1}$ for C and Si, respectively.  


These accretion rates are {\em a thousand times lower} than the value $\dot{M} = 1.7 \times 10^{-13} M_\odot$~yr$^{-1}$  measured from {\it XMM-Newton} X-ray spectra by \citet{Matranga.etal:12}.    The derived rates depend on the diffusion time scale and the mass of the convection zone.  While there are some
uncertainties for the deep convection zones that He-atmosphere
WDs have, calculations of the diffusion times in the thin convection 
zones of ``warm" H-atmosphere WDs are much more certain and have changed by a factor of 2 at most over the
past two decades---see, e.g., \citet{Paquette.etal:86}, \citet{Debes:06} and \citet{Koester:09}.  Moreover, the diffusion-based technique itself is mature and routinely applied to the analysis of WDs accreting from circumstellar material \citep[see, e.g., the review by][and references therein]{Jura:13}.
The difference we find in accretion rates is well beyond any systematic uncertainties in either method, even allowing for some deviation of the secondary metallicity from a solar composition.

\section{Accretion rate variations on QS~Vir}

We conclude that the accretion rate was completely different in the epochs of the {\it XMM-Newton} (2006) and {\it HST} (1999) observations separated by a 6.5 year interval.   While accretion rate variations are commonly observed on CVs and their underlying causes have been debated for years, in a detached system such variations are much more puzzling.  The variations on QS~Vir are, to the best of our knowledge, the first large accretion rate variations seen on a detached system.  We investigate below some possible mechanisms to account for these changes.  
 

\subsection{Captured coronal mass ejections?}

The accretion rate inferred from the HST spectrum obtained in 1999 is much too low to be caused by steady Roche Lobe overflow from the photosphere and is more consistent with wind accretion \citep[see the discussion of][]{Matranga.etal:12}.  It is an order of magnitude lower than the wind accretion rate inferred for similar pre-cataclysmic binaries binaries (\citealt{Debes:06,Tappert.etal:11,Pyrzas.etal:12}).  Moreover, it is difficult to conceive of a gradual mechanism that can change the accretion rate in a binary system by orders of magnitude over a maximum timescale of 6.5 years:  
\citet{Matranga.etal:12} noted that the timescale for accretion turn-on by angular momentum loss-driven gradual contact with the chromosphere or with spicule-like structures is $5\times 10^4$--$5\times 10^{5}$~yr.  Flare and associated coronal mass ejection (CME) or prominence activity on the red dwarf might inject mass within the white dwarf Roche surface on a stochastic basis.  The presence of material that might have originated from such a process was indeed inferred by \citet{Ribeiro.etal:10} based on Balmer line Doppler imaging and by \citet{Parsons.etal:11} from the detection of orbital phase-dependent H and Ca absorption along the line-of-sight towards the white dwarf.  Such a mechanism has also been invoked to explain apparent accretion rate variations during low states on magnetic CVs \citep[e.g.][]{Warren.etal:93,Pandel.Cordova:05}.  A CME explanation for accretion rate variations would imply that QS~Vir does not need to be in an unlikely, finely-balanced and fleeting evolutionary phase with Roche Lobe overflow through the upper chromosphere, as suggested by \citet{Matranga.etal:12}. 

Other than a moderate flare that was most likely associated with the M~dwarf, the eclipsed X-ray luminosity observed by  \citet{Matranga.etal:12} was essentially constant throughout the 110ks observation, indicating a constant accretion rate during this time.  
Integrated over the duration of the observation, this rate corresponds to about $10^{18}$g of material, or about a factor of 10 more mass than the largest solar coronal mass ejections compiled by \citet{Yashiro.Gopalswamy:09}.  This is a large, but not implausible, amount of mass, even if such an accretion event lasted a few days.  

The timescale for accretion of a CME event is likely quite short and occurs in a few orbits, taking perhaps of the order of a day.  A CME might have been associated with the X-ray flare seen by {\it XMM-Newton}, though since accretion was already ongoing at the start of the observation it is much more likely that a CME would have occurred earlier.  However, since we have only one observed flare, and 
again appealing to the \citet{Yashiro.Gopalswamy:09} compilation, we can estimate the mass that a CME associated with this flare might have contained.  The peak excess flare count rate above the M~dwarf quiescent rate was about 0.2~count~s$^{-1}$ in the EPIC-pn detector.  Using the scaling of EPIC-pn count rate to X-ray luminosity of \citet{Matranga.etal:12} and a flare duration of $\sim 5$ks, the X-ray fluence of the flare is $\sim 5\times 10^{32}$~erg.   \citet{Drake.etal:13b} find the mean CME ejected mass as a function of the associated solar flare fluence in the 1--8~\AA\ band to be $m_c \sim 0.032 {E_{1-8}}^{0.59}$.   For a flare plasma temperature of a few keV, the broad band 0.2--10~keV flux measured using EPIC is about 2--3 times higher than the 1--8~\AA\  flux, and the CME mass relation is approximately $m_c \sim 0.02 {E_{pn}}^{0.59}$, suggesting $m_c\sim 4\times 10^{17}$g.  

At face value, this mass is only a factor of two less than that required and perhaps sufficient to drive the observed accretion.  However, it does involve some extrapolation of the solar flare data, whose maximum fluence in the \citet{Yashiro.Gopalswamy:09}  sample would be about $1.5\times 10^{31}$~erg in the 0.2--10~keV band.  \citet{Drake.etal:13b} argue that the mean solar relation cannot be extrapolated to arbitrarily high energies and likely flattens for flare X-ray fluence above $10^{31}$~erg.   
Since most CMEs will likely not be accreted unless there is magnetic confinement within the binary system (see \S\ref{s:magswitch} below), the \citet{Matranga.etal:12} {\it XMM-Newton} observation would have been fortuitous to catch the object in the process of engulfing a CME unless the flaring rate were greater than seen in that 110~ks segment.  \citet{Matranga.etal:12} also noted that the ROSAT all-sky survey X-ray luminosity, corresponding to the epoch of late 1990, was $L_X=5\times 10^{29}$~erg~s$^{-1}$---similar to that of the {\it XMM-Newton} epoch.  If the accretion {\em was} due to stochastic CME events, these again must be relatively common, and more common than were seen by {\it XMM-Newton}.  The timescale for heavy element settling in Equation~\ref{e:mdot}, $\tau_X$, is only a few days for the white dwarf component of QS~Vir.  Accretion of a CME would cause a very rapid change to the HST UV metal lines, and traces of such a significant accretion episode would then be rapidly lost once the mass had been dissipated, as could have been the case during the 1999 epoch.

Stochastic mass accretion episodes are, then, not inconsistent with the observations, though the likelihood of the 2006 accretion rate being dominated by a CME is perhaps diminished by the presence of only a single flare in the {\it XMM-Newton} light curve.  We are therefore drawn to examine possible alternative explanations for the different accretion rates seen at different epochs.  

\subsection{Perturbation of the orbit due to the tertiary component?}

The distant tertiary component of the system inferred by \citet{Parsons.etal:10} appears to be in a highly elliptical orbit with a period of 14~yr.   The perigee passage in their inferred orbit is near the beginning of 2006---essentially the same epoch as the {\it XMM-Newton} observation when a ``high'' accretion rate was observed.   Instead, the {\it HST} observation indicating a very low rate was obtained about 6.5 years earlier, close to apogee.   The {\it ROSAT} data obtained about 8 years before that and indicating a high rate, again correspond fairly closely to the inferred perigee.  This is suggestive of a scenario in which the modulation in accretion rate is caused by perturbation of the central binary orbit by the tertiary component.  

The gravitational effect of a third body on a two-body system
resembles, instantaneously, that of a tidal effect.  Thus, we expect the orbital separation of the binary system to change with the phase of the third body on its eccentric
orbit.  Such a  perturbation would render the central binary orbit slightly elliptical.  At apogee, the third body has relatively little influence on the central system compared with the situation at perigee.  At perigee we might expect the orbit to be more elliptical---stretched along the position vector to the tertiary component and shrunk along the axis perpendicular to it.  If a big enough effect, such shrinkage of the separation might precipitate enhanced accretion, turning it on and off with the tertiary body orbital phase.  

Assuming the third body is not affected by the motion of the central binary system,
we can use Perturbation Theory to estimate the magnitude of the effect (see Appendix~A), and 
find that perturbations are smaller than
$10^{-7} R_{sun}$.  \citet{Matranga.etal:12} argue that the low rate accretion likely occurs from the chromosphere.  For the third body perturbations to work as an accretion switch, the orbital separation must then change by an amount commensurate with the chromospheric gas scale height.  For gas with temperature $T_{ch}$, this is $h_{ch}=kT_{ch}{R_2}^2/\mu Gm_2$, and for a chromospheric temperature of $10^4$~K, the ratio of the scale height to the secondary star radius, $R_2$, is $h_{ch}/R_2=4.3\times 10^{-4}$.  The orbital perturbations are therefore orders of magnitude too small to affect the accretion rate, and perturbations to the position of the L1 point within the orbit are similarly small, with changes less than $10^{-7}R_2$.

If the period variations are interpreted only in terms of orbital separation changes, neglecting angular momentum issues for simplicity, differentiation of Kepler's third law implies $\Delta a/R_2=2\Delta P/P$, where the stellar radius $R_2$ is about 1/3 the orbital separation, $a$ \citep[e.g.][]{O'Donoghue.etal:03,Ribeiro.etal:10}.  For a change of eclipse times $O-C$ of 150s over 4 years from 2002-2006 in Figure~10 of \citet{Parsons.etal:10}, $\Delta a/R_2\sim 2\times 10^{-6}$--two orders of magnitude smaller than the chromospheric scale height. 


\subsection{A magnetic accretion switch?}
\label{s:magswitch}

There are at least three ways in which magnetic fields might affect the accretion rate on sufficiently short timescales to explain the ``on'' and ``off" states seen in different years.  One is the \citet{Applegate:92} mechanism, originally proposed to explain observed period changes in close binary stars, somewhat like those seen for QS~Vir itself by \citet{O'Donoghue.etal:03}, \citet{Qian.etal:10} and \citet{Parsons.etal:10}; the second is the influence of the interaction between the large-scale field of the M dwarf and white dwarf that can affect the rate of accretion of the M dwarf magnetically-driven wind and recently studied by \citet{Cohen.etal:12}.  The third is the occurrance of starspots at the L1 point that \citet{Livio.Pringle:94} proposed in order to explain CV ``low states'' of the VY~Scl class of CVs that experience abrupt drops in accretion rate.  Similarly, the irregular changes in the accretion rate observed among CVs with strongly magnetic white dwarfs (polars) can be described by spotted donor stars \citep{Hessman.etal:00}.

Of these mechanisms, we can readily dismiss that of \citet{Applegate:92} on the grounds that any effects are too small to affect the accretion rate.  There are two relevant aspects of the redistribution of angular momentum within the secondary that is the basis of the mechansim: orbital period modulation and the associated change in the semi-major axis; and changes in oblateness of the secondary.  The former effect is of the order of $\Delta a/a=\Delta P/P$, and, following \citet{Parsons.etal:10},  energy requirements of period changes imply $\Delta P/P\la 10^{-7}$.  Similarly, we find that a distortion of the order of $\Delta R/R \sim 10^{-4}$ requires orders of magnitude greater energy than can be supplied by the stellar luminosity, unless only an unrealistically thin shell of mass $\sim 10^{-4}M_2$ is involved.

\citet{Cohen.etal:12} have studied the magnetic interaction between a synchronously rotating M dwarf and white dwarf in a detached binary.   They found that for some values of the respective white dwarf and M dwarf magnetic fields and alignments, an efficient syphoning of coronal plasma from the inward facing M dwarf hemisphere occurs.  The wind accretion rates were found to depend on the alignment of the two fields and, consequently, that wind accretion should be modulated by the magnetic cycle on the M~dwarf.   One particular configuration modelled by \citet{Cohen.etal:12} showed a cyclic change in accretion rate by a factor of 6.  While this is insufficient to explain the orders of magnitude drop in accretion implied by HST observations, the simulations showed that accretion rates are quite sensitive to the orbital and magnetic field parameters adopted.  Further simulations exploring the influence of the orbital separation and magnetic field strengths on the magnitude of the accretion rate cyclic modulation would be of interested in this context.    Whether the M dwarf in QS~Vir experiences magnetic cycles akin to that of the Sun remains an open question.  \citet{Savanov:12} finds that M~dwarf cycles, including those for fairly rapid rotators, tend to adhere to periods well-behaved in the $\log P_{rot}$~vs.~$\log P_{cyc}/P_{rot}$ plane, in rough accord with dynamo theory as found for earlier spectral types \citep[e.g.][]{Baliunas.etal:96,Olah.etal:09}.   Based on \citet{Savanov:12}, we would expect a cyclic period for QS~Vir of the order of $\log P_{cyc}/P_{rot} \sim 3.5$, or $P_{cyc}\sim 470$d, give or take a factor of 2--3.   The \citet{Parsons.etal:10} $O-C$ diagram for eclipse times in QS~Vir exhibits residual wobble with a timescale of several years which is perhaps consistent with this---those authors indeed note that this much smaller effect on the period is energetically consistent with the \citet{Applegate:92} mechanism.

Perhaps more difficult for a magnetic cycle accretion switch for QS~Vir is the wind-driven mass loss rate that the {\it XMM-Newton} and ROSAT X-ray fluxes would demand.  \citet{Matranga.etal:12} argue that, for an accretion efficiency of the order of 10\%, the required $\sim 2\times 10^{-12}M_\odot$~yr$^{-1}$ is perhaps high in comparison to the best current estimates of M~dwarf wind mass loss rates based on observations of pre-polar and other pre-CVs that lie in the range $10^{-13}$--$10^{-15} M_\odot$~yr$^{-1}$ 
\citep[e.g.][]{Schwope.etal:02,Schmidt.etal:05,Schmidt.etal:07,Vogel.etal:11,Debes:06,Tappert.etal:11}.   This problem would be alleviated if the M~dwarf and WD separation and magnetic field strengths allowed the collection of nearly all the wind of the former, as occured in \citet{Cohen.etal:12}  Model F.

The mechanism of \citet{Livio.Pringle:94}  invoked to explain low accretion states of the VY~Scl stars posits that magnetic star spots located over the L1 point can inhibit accretion through lowering of the gas scale height.  This is achieved because of the lower photospheric temperatures of star spot regions in which magnetic pressure is significant compared with the ambient gas pressure.    \citet{Ritter:88} also noted that flow through the L1 point could be temporarily inhibited by closed magnetic field lines at L1, provided such fields were of kG strength.   Since both direct magnetic inhibition and reduction in the gas scale height require localised star spots with strong magnetic fields, these two mechanisms will be difficult to distinguish observationally.  
Nevertheless, with stochastic accretion of CMEs and magnetic modulation of wind accretion both appearing to face at least minor difficulties, these types of magnetic modulation could also be possible if QS~Vir were indeed accreting from the high chromosphere as suggested by \citet{Matranga.etal:12}.  Rigorous testing of this model would require Doppler imaging of the M~dwarf to locate star spots, combined with simultaneous observations at UV or X-ray wavelengths to monitor the accretion rate.  The former two mechanisms---CMEs and magnetic cycling switching---are more simple to test.  Following the accretion rate on timescales of weeks, to probe for CME-driven modulation, to a year or two to, to probe magnetic cycle switching, could be achieved with HST~UV spectroscopy.  X-ray observations would also be greatly beneficial to probe flaring, and indirectly, CME behavior.

Finally, despite the implausibility of the \citet{Applegate:92} mechanism being responsible for the accretion rate variations, it is still tempting to draw a connection between these variations and orbital period variations that have a period similar to that expected from magnetic cycles.   



\section{Is QS Vir just hibernating?}

The seminal study of QS~Vir by \citet{O'Donoghue.etal:03} suggested the binary could be a ``hibernating'' cataclysmic variable rather than a pre-contact system.  They based this suggestion on inference from line profiles that the white dwarf appeared to be rotating rapidly---a sign of accretion-driven spin-up---and that the H$\alpha$ emission line showed evidence of absorption by an accretion stream.  However, \citet{Parsons.etal:11} found the Mg~{\sc II} 4481\AA\ line to be narrow and consistent with a low rotation rate, while both \citet{Ribeiro.etal:10} and \citet{Parsons.etal:11} found the H$\alpha$ profile and its variations inconsistent with an accretion stream.  We note here, however, that the white dwarf in the archetypal dwarf nova U~Gem is a rather slow rotator \citep{Sion.etal:94,Long.etal:06}, and a low rotation rate is therefore not a definitive test of absence of prior accretion episodes.  A hibernating state for QS~Vir should therefore still be entertained.

Hibernation might set in as a result of either mass loss and associated widening of the binary separation following a nova explosion \citep{Shara.etal:86}, or irradiation-driven mass transfer cycles cycles \citep[e.g.][]{Podsiadlowski:91,Buning.Ritter:04}.  QS~Vir could conceivably fit into either category.  \citet{Buning.Ritter:04} find that CVs close to the upper edge of the period gap can undergo cycles if angular momentum loss rates are low compared with the prescription of \citet{Verbunt.Zwaan:81} and similar to the rates favoured by \citet[e.g.][]{Sills.etal:00} and \citet{Andronov.etal:03}.  Some evidence for a lower angular momentum loss rate just above the CV period gap is reviewed by \citet{Knigge.etal:11}.  More direct support for irradiation-driven cycles comes from very high mean accretion rates typically inferred for CVs in the 3-4~hr period range \citep[e.g.][]{Townsley.Gansicke:09}---higher than would be the case if driven purely by angular momentum loss.  In either case, QS~Vir could represent the extreme low end of an $\dot{M}$ spectrum resulting from very large amplitude mass transfer cycles.  The present accretion rate variations could not easily be interpretted simply as a result of going into hibernation, however,  because relaxation timescales are orders of magnitude larger than a year \citep[e.g.][]{Knigge.etal:11}.

\section{Conclusions}

Examination of the photospheric abundances of QS~Vir derived by \citet{O'Donoghue.etal:03} based on 1999 HST STIS observations indicates an accretion rate of $\sim 10^{-16}M_\odot$~yr$^{-1}$, which is a thousand times lower than the already very low accretion rate observed by \cite{Matranga.etal:12} using {\it XMM-Newton} in 2006 and inferred by the same authors from the ROSAT all-sky survey in 1991.   This is the first time large accretion rate variations have been seen in a detached close binary.

It is tempting to ascribe the change in accretion rate to the influence of a putative third body in the QS~Vir binary system, but we find gravitational perturbations to the central binary caused by such a body are orders of magnitude too small.  Orbital and oblateness perturbations that might be ascribed to the \citet{Applegate:92} mechanism are also much too small.  The accretion rate could be unsteady if fed by coronal mass ejections.  This has the advantage of not requiring QS~Vir to be in such a finely-poised evolutionary state so as to be accreting at a low rate through the chromosphere, but would require more flaring than observed by {\it XMM-Newton}.  We are also led to suggest a magnetic accretion switch acts on QS~Vir.   One  mechanism is star spots at the L1 point, originally suggested to explain the low states in the VY~Scl class of CVs, that could modulate accretion through L1 overflow by lowering the pressure scale height \citep{Livio.Pringle:94} or by direct magnetic confinement \citep{Ritter:88}.  This mechanism again requires a finely-balanced, tenuous chromospheric gas supply.  We consider more likely an M~dwarf secondary magnetic cycle polarity switch that can strongly influence the wind accretion rate \citep{Cohen.etal:12}.  The latter probably requires a wind-driven mass loss rate of the order of $10^{-12} M_\odot$~yr$^{-1}$, which is perhaps rather high.  These mechanisms and stochastic accretion events might be readily distinguished by UV and X-ray monitoring combined with Doppler imaging of the secondary component.  

Wind or CME-dominated accretion would mean that UV and X-ray observations of pre-CVs could provide a powerful probe of the magnetospheres, mass loss, and angular momentum loss of CV-like secondary stars.  



\section*{Acknowledgments}

JJD was supported by NASA contract NAS8-03060 to the {\it
Chandra X-ray Center} (CXC) and
thanks the CXC director, Harvey Tananbaum, and the CXC science team
for advice and support.  DT was supported by JSPS Fellowship.  
Finally, we thank G.~Schmidt for comments on Matranga et al.\ (2012) that provided the initial impetus for this further study.

\bibliographystyle{mn2e}

\begin{thebibliography}{}

\bibitem[\protect\citeauthoryear{{Almeida} \& {Jablonski}}{{Almeida} \&
  {Jablonski}}{2011}]{Almeida.Jablonski:11}
{Almeida} L.~A.,  {Jablonski} F.,  2011, in {Sozzetti} A.,  {Lattanzi} M.~G.,
  {Boss} A.~P.,  eds, IAU Symposium Vol.~276 of IAU Symposium, {Two bodies with
  high eccentricity around the cataclysmic variable QS Vir}.
pp 495--496

\bibitem[\protect\citeauthoryear{{Andronov}, {Pinsonneault} \&
  {Sills}}{{Andronov} et~al.}{2003}]{Andronov.etal:03}
{Andronov} N.,  {Pinsonneault} M.,    {Sills} A.,  2003, \apj, 582, 358

\bibitem[\protect\citeauthoryear{{Applegate}}{{Applegate}}{1992}]{Applegate:92}
{Applegate} J.~H.,  1992, \apj, 385, 621

\bibitem[\protect\citeauthoryear{{Baliunas}, {Nesme-Ribes}, {Sokoloff} \&
  {Soon}}{{Baliunas} et~al.}{1996}]{Baliunas.etal:96}
{Baliunas} S.~L.,  {Nesme-Ribes} E.,  {Sokoloff} D.,    {Soon} W.~H.,  1996,
  \apj, 460, 848

\bibitem[\protect\citeauthoryear{{Bil{\'{\i}}kov{\'a}}, {Chu}, {Gruendl} \&
  {Maddox}}{{Bil{\'{\i}}kov{\'a}} et~al.}{2010}]{Bilikova.etal:10}
{Bil{\'{\i}}kov{\'a}} J.,  {Chu} Y.,  {Gruendl} R.~A.,    {Maddox} L.~A.,
  2010, \aj, 140, 1433

\bibitem[\protect\citeauthoryear{{Brown} \& {Shook}}{{Brown} \&
  {Shook}}{1933}]{Brown.Shook:33}
{Brown} E.~W.,  {Shook} C.~A.,  1933, {Planetary Theory}.
Cambridge University press, Cambridge, England

\bibitem[\protect\citeauthoryear{{B{\"u}ning} \& {Ritter}}{{B{\"u}ning} \&
  {Ritter}}{2004}]{Buning.Ritter:04}
{B{\"u}ning} A.,  {Ritter} H.,  2004, \aap, 423, 281

\bibitem[\protect\citeauthoryear{{Cohen}, {Drake} \& {Kashyap}}{{Cohen}
  et~al.}{2012}]{Cohen.etal:12}
{Cohen} O.,  {Drake} J.~J.,    {Kashyap} V.~L.,  2012, ArXiv e-prints

\bibitem[\protect\citeauthoryear{{Debes}}{{Debes}}{2006}]{Debes:06}
{Debes} J.~H.,  2006, \apj, 652, 636

\bibitem[\protect\citeauthoryear{{Drake}, {Cohen}, {Yashiro} \&
  {Gopalswamy}}{{Drake} et~al.}{2013}]{Drake.etal:13b}
{Drake} J.~J.,  {Cohen} O.,  {Yashiro} S.,    {Gopalswamy} N.,  2013, \apj,
  764, 170

\bibitem[\protect\citeauthoryear{{Dupuis}, {Fontaine} \& {Wesemael}}{{Dupuis}
  et~al.}{1993}]{Dupuis.etal:93}
{Dupuis} J.,  {Fontaine} G.,    {Wesemael} F.,  1993, \apjs, 87, 345

\bibitem[\protect\citeauthoryear{{Hessman}, {G{\"a}nsicke} \&
  {Mattei}}{{Hessman} et~al.}{2000}]{Hessman.etal:00}
{Hessman} F.~V.,  {G{\"a}nsicke} B.~T.,    {Mattei} J.~A.,  2000, \aap, 361,
  952

\bibitem[\protect\citeauthoryear{{Horner}, {Wittenmyer}, {Hinse}, {Marshall},
  {Mustill} \& {Tinney}}{{Horner} et~al.}{2013}]{Horner.etal:13}
{Horner} J.,  {Wittenmyer} R.~A.,  {Hinse} T.~C.,  {Marshall} J.~P.,  {Mustill}
  A.~J.,    {Tinney} C.~G.,  2013, \mnras

\bibitem[\protect\citeauthoryear{{Jura}}{{Jura}}{2013}]{Jura:13}
{Jura} M.,  2013, ArXiv e-prints

\bibitem[\protect\citeauthoryear{{Kawaler}}{{Kawaler}}{1988}]{Kawaler:88}
{Kawaler} S.~D.,  1988, \apj, 333, 236

\bibitem[\protect\citeauthoryear{{Kawka}, {Vennes}, {Koch} \&
  {Williams}}{{Kawka} et~al.}{2002}]{Kawka.etal:02}
{Kawka} A.,  {Vennes} S.,  {Koch} R.,    {Williams} A.,  2002, \aj, 124, 2853

\bibitem[\protect\citeauthoryear{{Kilkenny}, {Koen}, {O'Donoghue} \&
  {Stobie}}{{Kilkenny} et~al.}{1997}]{Kilkenny.etal:97}
{Kilkenny} D.,  {Koen} C.,  {O'Donoghue} D.,    {Stobie} R.~S.,  1997, \mnras,
  285, 640

\bibitem[\protect\citeauthoryear{{Knigge}, {Baraffe} \& {Patterson}}{{Knigge}
  et~al.}{2011}]{Knigge.etal:11}
{Knigge} C.,  {Baraffe} I.,    {Patterson} J.,  2011, \apjs, 194, 28

\bibitem[\protect\citeauthoryear{{Koester}}{{Koester}}{2009}]{Koester:09}
{Koester} D.,  2009, \aap, 498, 517

\bibitem[\protect\citeauthoryear{{Kopal}}{{Kopal}}{1978}]{Kopal:78}
{Kopal} Z.,  1978, {Dynamics of close binary systems}.
Vol.~68 of Astrophysics and Space Science Library

\bibitem[\protect\citeauthoryear{{Livio} \& {Pringle}}{{Livio} \&
  {Pringle}}{1994}]{Livio.Pringle:94}
{Livio} M.,  {Pringle} J.~E.,  1994, \apj, 427, 956

\bibitem[\protect\citeauthoryear{{Long}, {Brammer} \& {Froning}}{{Long}
  et~al.}{2006}]{Long.etal:06}
{Long} K.~S.,  {Brammer} G.,    {Froning} C.~S.,  2006, \apj, 648, 541

\bibitem[\protect\citeauthoryear{{Matranga}, {Drake}, {Kashyap} \&
  {Steeghs}}{{Matranga} et~al.}{2012}]{Matranga.etal:12}
{Matranga} M.,  {Drake} J.~J.,  {Kashyap} V.,    {Steeghs} D.,  2012, \apj,
  747, 132

\bibitem[\protect\citeauthoryear{{Mestel}}{{Mestel}}{1968}]{Mestel:68}
{Mestel} L.,  1968, \mnras, 138, 359

\bibitem[\protect\citeauthoryear{{O'Donoghue}, {Koen}, {Kilkenny}, {Stobie},
  {Koester}, {Bessell}, {Hambly} \& {MacGillivray}}{{O'Donoghue}
  et~al.}{2003}]{O'Donoghue.etal:03}
{O'Donoghue} D.,  {Koen} C.,  {Kilkenny} D.,  {Stobie} R.~S.,  {Koester} D.,
  {Bessell} M.~S.,  {Hambly} N.,    {MacGillivray} H.,  2003, \mnras, 345, 506

\bibitem[\protect\citeauthoryear{{Ol{\'a}h}, {Koll{\'a}th}, {Granzer},
  {Strassmeier}, {Lanza}, {J{\"a}rvinen}, {Korhonen}, {Baliunas}, {Soon},
  {Messina} \& {Cutispoto}}{{Ol{\'a}h} et~al.}{2009}]{Olah.etal:09}
{Ol{\'a}h} K.,  {Koll{\'a}th} Z.,  {Granzer} T.,  {Strassmeier} K.~G.,  {Lanza}
  A.~F.,  {J{\"a}rvinen} S.,  {Korhonen} H.,  {Baliunas} S.~L.,  {Soon} W.,
  {Messina} S.,    {Cutispoto} G.,  2009, \aap, 501, 703

\bibitem[\protect\citeauthoryear{{Paczynski}}{{Paczynski}}{1976}]{Paczynski:76}
{Paczynski} B.,  1976, in {P.~Eggleton, S.~Mitton, \& J.~Whelan} ed., Structure
  and Evolution of Close Binary Systems Vol.~73 of IAU Symposium, {Common
  Envelope Binaries}.
pp 75--+

\bibitem[\protect\citeauthoryear{{Pandel} \& {C{\'o}rdova}}{{Pandel} \&
  {C{\'o}rdova}}{2005}]{Pandel.Cordova:05}
{Pandel} D.,  {C{\'o}rdova} F.~A.,  2005, \apj, 620, 416

\bibitem[\protect\citeauthoryear{{Paquette}, {Pelletier}, {Fontaine} \&
  {Michaud}}{{Paquette} et~al.}{1986}]{Paquette.etal:86}
{Paquette} C.,  {Pelletier} C.,  {Fontaine} G.,    {Michaud} G.,  1986, \apjs,
  61, 197

\bibitem[\protect\citeauthoryear{{Parsons}, {Marsh}, {Copperwheat}, {Dhillon},
  {Littlefair}, {Hickman}, {Maxted}, {G{\"a}nsicke}, {Unda-Sanzana}, {Colque},
  {Barraza}, {S{\'a}nchez} \& {Monard}}{{Parsons}
  et~al.}{2010}]{Parsons.etal:10}
{Parsons} S.~G.,  {Marsh} T.~R.,  {Copperwheat} C.~M.,  {Dhillon} V.~S.,
  {Littlefair} S.~P.,  {Hickman} R.~D.~G.,  {Maxted} P.~F.~L.,  {G{\"a}nsicke}
  B.~T.,  {Unda-Sanzana} E.,  {Colque} J.~P.,  {Barraza} N.,  {S{\'a}nchez} N.,
     {Monard} L.~A.~G.,  2010, \mnras, 407, 2362

\bibitem[\protect\citeauthoryear{{Parsons}, {Marsh}, {G{\"a}nsicke} \&
  {Tappert}}{{Parsons} et~al.}{2011}]{Parsons.etal:11}
{Parsons} S.~G.,  {Marsh} T.~R.,  {G{\"a}nsicke} B.~T.,    {Tappert} C.,  2011,
  \mnras, 412, 2563

\bibitem[\protect\citeauthoryear{{Podsiadlowski}}{{Podsiadlowski}}{1991}]{Pods%
iadlowski:91}
{Podsiadlowski} P.,  1991, \nat, 350, 136

\bibitem[\protect\citeauthoryear{{Pyrzas}, {G{\"a}nsicke}, {Brady}, {Parsons},
  {Marsh}, {Koester}, {Breedt}, {Copperwheat}, {Nebot G{\'o}mez-Mor{\'a}n},
  {Rebassa-Mansergas}, {Schreiber} \& {Zorotovic}}{{Pyrzas}
  et~al.}{2012}]{Pyrzas.etal:12}
{Pyrzas} S.,  {G{\"a}nsicke} B.~T.,  {Brady} S.,  {Parsons} S.~G.,  {Marsh}
  T.~R.,  {Koester} D.,  {Breedt} E.,  {Copperwheat} C.~M.,  {Nebot
  G{\'o}mez-Mor{\'a}n} A.,  {Rebassa-Mansergas} A.,  {Schreiber} M.~R.,
  {Zorotovic} M.,  2012, \mnras, 419, 817

\bibitem[\protect\citeauthoryear{{Qian}, {Liao}, {Zhu}, {Dai}, {Liu}, {He},
  {Zhao} \& {Li}}{{Qian} et~al.}{2010}]{Qian.etal:10}
{Qian} S.,  {Liao} W.,  {Zhu} L.,  {Dai} Z.,  {Liu} L.,  {He} J.,  {Zhao} E.,
   {Li} L.,  2010, \mnras, 401, L34

\bibitem[\protect\citeauthoryear{{Ribeiro}, {Kafka}, {Baptista} \&
  {Tappert}}{{Ribeiro} et~al.}{2010}]{Ribeiro.etal:10}
{Ribeiro} T.,  {Kafka} S.,  {Baptista} R.,    {Tappert} C.,  2010, \aj, 139,
  1106

\bibitem[\protect\citeauthoryear{{Ritter}}{{Ritter}}{1988}]{Ritter:88}
{Ritter} H.,  1988, \aap, 202, 93

\bibitem[\protect\citeauthoryear{{Savanov}}{{Savanov}}{2012}]{Savanov:12}
{Savanov} I.~S.,  2012, Astronomy Reports, 56, 716

\bibitem[\protect\citeauthoryear{{Schmidt}, {Szkody}, {Henden}, {Anderson},
  {Lamb}, {Margon} \& {Schneider}}{{Schmidt} et~al.}{2007}]{Schmidt.etal:07}
{Schmidt} G.~D.,  {Szkody} P.,  {Henden} A.,  {Anderson} S.~F.,  {Lamb} D.~Q.,
  {Margon} B.,    {Schneider} D.~P.,  2007, \apj, 654, 521

\bibitem[\protect\citeauthoryear{{Schmidt}, {Szkody}, {Vanlandingham},
  {Anderson}, {Barentine}, {Brewington}, {Hall}, {Harvanek}, {Kleinman},
  {Krzesinski}, {Long}, {Margon}, {Neilsen} Jr., {Newman}, {Nitta}, {Schneider}
  \& {Snedden}}{{Schmidt} et~al.}{2005}]{Schmidt.etal:05}
{Schmidt} G.~D.,  {Szkody} P.,  {Vanlandingham} K.~M.,  {Anderson} S.~F.,
  {Barentine} J.~C.,  {Brewington} H.~J.,  {Hall} P.~B.,  {Harvanek} M.,
  {Kleinman} S.~J.,  {Krzesinski} J.,  {Long} D.,  {Margon} B.,  {Neilsen} Jr.
  E.~H.,  {Newman} P.~R.,  {Nitta} A.,  {Schneider} D.~P.,    {Snedden} S.~A.,
  2005, \apj, 630, 1037

\bibitem[\protect\citeauthoryear{{Schwope}, {Brunner}, {Hambaryan} \&
  {Schwarz}}{{Schwope} et~al.}{2002}]{Schwope.etal:02}
{Schwope} A.~D.,  {Brunner} H.,  {Hambaryan} V.,    {Schwarz} R.,  2002, in
  {B.~T.~G{\"a}nsicke, K.~Beuermann, \& K.~Reinsch} ed., The Physics of
  Cataclysmic Variables and Related Objects Vol.~261 of Astronomical Society of
  the Pacific Conference Series, {LARPs -- Low-accretion rate polars}.
p.~102

\bibitem[\protect\citeauthoryear{{Shara}, {Livio}, {Moffat} \& {Orio}}{{Shara}
  et~al.}{1986}]{Shara.etal:86}
{Shara} M.~M.,  {Livio} M.,  {Moffat} A.~F.~J.,    {Orio} M.,  1986, \apj, 311,
  163

\bibitem[\protect\citeauthoryear{{Sills}, {Pinsonneault} \& {Terndrup}}{{Sills}
  et~al.}{2000}]{Sills.etal:00}
{Sills} A.,  {Pinsonneault} M.~H.,    {Terndrup} D.~M.,  2000, \apj, 534, 335

\bibitem[\protect\citeauthoryear{{Sion}, {Long}, {Szkody} \& {Huang}}{{Sion}
  et~al.}{1994}]{Sion.etal:94}
{Sion} E.~M.,  {Long} K.~S.,  {Szkody} P.,    {Huang} M.,  1994, \apjl, 430,
  L53

\bibitem[\protect\citeauthoryear{{Tappert}, {G{\"a}nsicke}, {Schmidtobreick} \&
  {Ribeiro}}{{Tappert} et~al.}{2011}]{Tappert.etal:11}
{Tappert} C.,  {G{\"a}nsicke} B.~T.,  {Schmidtobreick} L.,    {Ribeiro} T.,
  2011, \aap, 532, A129

\bibitem[\protect\citeauthoryear{{Townsley} \& {G{\"a}nsicke}}{{Townsley} \&
  {G{\"a}nsicke}}{2009}]{Townsley.Gansicke:09}
{Townsley} D.~M.,  {G{\"a}nsicke} B.~T.,  2009, \apj, 693, 1007

\bibitem[\protect\citeauthoryear{{Verbunt} \& {Zwaan}}{{Verbunt} \&
  {Zwaan}}{1981}]{Verbunt.Zwaan:81}
{Verbunt} F.,  {Zwaan} C.,  1981, \aap, 100, L7

\bibitem[\protect\citeauthoryear{{Vogel}, {Schwope} \& {Schwarz}}{{Vogel}
  et~al.}{2011}]{Vogel.etal:11}
{Vogel} J.,  {Schwope} A.~D.,    {Schwarz} R.,  2011, \aap, 530, A117+

\bibitem[\protect\citeauthoryear{{Warren}, {Vallerga}, {Mauche}, {Mukai} \&
  {Siegmund}}{{Warren} et~al.}{1993}]{Warren.etal:93}
{Warren} J.~K.,  {Vallerga} J.~V.,  {Mauche} C.~W.,  {Mukai} K.,    {Siegmund}
  O.~H.~W.,  1993, \apjl, 414, L69

\bibitem[\protect\citeauthoryear{{Weber} \& {Davis} Jr.}{{Weber} \&
  {Davis}}{1967}]{Weber.Davis:67}
{Weber} E.~J.,  {Davis} Jr. L.,  1967, \apj, 148, 217

\bibitem[\protect\citeauthoryear{{Yashiro} \& {Gopalswamy}}{{Yashiro} \&
  {Gopalswamy}}{2009}]{Yashiro.Gopalswamy:09}
{Yashiro} S.,  {Gopalswamy} N.,  2009, in {Gopalswamy} N.,  {Webb} D.~F.,  eds,
  IAU Symposium Vol.~257 of IAU Symposium, {Statistical relationship between
  solar flares and coronal mass ejections}.
pp 233--243

\bibitem[\protect\citeauthoryear{{Zorotovic} \& {Schreiber}}{{Zorotovic} \&
  {Schreiber}}{2013}]{Zorotovic.Schreiber:13}
{Zorotovic} M.,  {Schreiber} M.~R.,  2013, \aap, 549, A95

\end{thebibliography}

\appendix

\section{Gravitational perturbation to a compact binary orbit due to a more distant orbiting third body} 

For the case in which the third body is not affected by the motion of the central binary system,  
Perturbation Theory can be used to estimate the changes to the central orbit.
The only orbital elements that affect the separation of the components 
are the semi-major axes $a$ and the eccentricity $e$.
The effect of the tertiary body on the orbital elements of the
binary are encoded in the Disturbing Function \citet{Brown.Shook:33}, which
represents the gravitational potential of the perturbation and takes
the form 
\begin{equation*}
S= G \, m_3 \frac{r^2}{{r'}^3} \sum
\frac{(m_1)^j-(-m_2)^j}{(m_1+m_2)^j} (\frac{r}{r'})^{j-1}
P_{j+1}(\sigma)+ G \, \frac{(m_1+m_2)}{2a}
\end{equation*}
Here, $G$ denotes the gravitational constant; $r$ and $r'$ stand for
the radius vectors of the binary and third body orbits; $a$ for the
binary semi-major axis; $m_1$ and
$m_2$ for the masses of the binary components; $m_3$ for the mass of the 
tertiary body; and $P_j(\sigma)$ are the Legendre polynomials
of $\sigma$, the cosine of the inclination $i$ of the wide orbit. 
By changing to Delaunay variables ($L$, $G$,
$H$, $l$, $g$ and $h$), the variational
equations take their canonical form. The Delaunay individual elements
can then be obtained by partially
differentiating $I=\int S dt$ with respect to the desired
element. The actual orbital elements $a$,
$e$, $i$, longitude of the pericenter $\omega$, mean daily motion $n$,
time of the periastron passage $T$, and the longitude of the node $\Omega$ are easily obtained from the
Delaunay variables as follows:
\begin{align*}
a &= \sqrt{L/n},&   n(t-T) &= l \\
e &= \sqrt{1-(G/L)^2}, &  w &= g\\
i &= cos^{-1}(H/G); &  \Omega &= h.\\
\end{align*}

Following \citet{Kopal:78}, we separate the effect
into two categories: short-range perturbations, corresponding to one period of
the orbit of the binary system, and assuming the third body is fixed at the perigee;
and long-range perturbations, corresponding to one period of the orbit of the third
body, and for which all short-periodic terms of the
disturbing function are omitted, and its integrals are averaged over the
binary period.   For the short-range perturbations, the variational equation for the
eccentricity takes the form
\begin{equation*}
\frac{\delta e}{2 \pi \sqrt{1-e^2}} = 
-\frac{15}{2} \kappa_{1} M N e+ \frac{5}{8} \kappa_{2} [(1-e^2) P'_{3}(M)+ (1+6 e^2) P'_{3}(N)-3(1-e^2) ] M
\end{equation*}
where
\begin{align*}
M &= - \sin{\omega} \cos{u'} + \cos{\omega} \sin{u'} \cos{i},\\ 
N &=  \cos{\omega} \cos{u'} + \sin{\omega} \sin{u'} \cos{i},\\
\kappa_{1} &= \frac{m_3}{m_1+m_2} \big( \frac{a}{r'}\big)^3,\\
\kappa_{2} &= \frac{m_3(m_1-m_2)}{(m_1+m_2)^2} \big( \frac{a}{r'}\big)^4,\\
u &=\omega + \nu,  \  \  \text{and} \ \ q = \cos{i}.
\end{align*}
As usual, $ \nu$ is the true anomaly and the primed quantities refer to the tertiary body.

For the long-range perturbations, if the tertiary component and the binary orbits happen to be coplanar, the secular variation of the eccentricity becomes
\begin{align*}
\delta e &= -\frac{4}{5} \kappa'_{2} e' (1-e^2)(4+3 e^2) \sin{(\omega' - \omega)}.
\intertext{For the case where the central orbit is circular but they are not coplanar, the respective perturbation assumes the form}
\delta e &= -\frac{1}{5} \kappa'_{2} e' \{(5 q^2+11) \sin{\omega} \cos{\omega'}- q (15 q^2+1) \cos{\omega}\sin{\omega' }\}
\end{align*}
where
\begin{align*}
\kappa'_{1} &= \frac{3}{4}\frac{\pi \tilde{G} m_3}{n n' {a'}^3 \sqrt{(1-e^2)(1-{e'}^2)^3}},\\
\intertext{and}
\kappa'_{2} &= \frac{25}{8}\Big(\frac{m_1-m_2}{m_1+m_2}\Big) \frac{a \kappa'_{1}}{a'(1-{e'}^2)}.\\
\end{align*}
The period of the binary is much shorter than that of the perturbing body.  Therefore, the position of the tertiary in its orbit is only relevant when computing the distance to the center of mass of the binary and not their relative orientation. For this reason, we expect the effect to be maximum when the third body is at its perigee and minimum at its apogee.  We assume the perturbation to be negligible at apogee.  By this we are setting a higher boundary on the difference between the maximum and minimum effect. 

The variation in the distance between the binary components resulting from the short-range perturbations, $\delta e \, a$, is of order $10^{-10} R_{sun}$ for one orbital period of the binary. Multiplying this effect by the number of binary periods that occur in the time it takes the tertiary component to go from $\nu' = -\pi/2 $ to $\nu'=\pi/2$ we find the perturbation to be of the order of $10^{-7} R_{sun}$. This is an overestimation since we are assuming the perturbing body to be fixed at perigee during all this time.   The long-range perturbations are of the order of $10^{-14}R_{sun}$ for the period of the wide orbit.  


\end{document}